\documentclass[twocolumn]{aastex631}

\received{}
\revised{}
\accepted{}

\submitjournal{The Astrophysical Journal}

\shorttitle{Reacceleration of Nonthermal Electrons at ICM Shocks}
\shortauthors{Ha et al.}

\begin{document}

\title{Electron Preacceleration at Weak Quasi-perpendicular Intracluster Shocks: Effects of Preexisting Nonthermal Electrons}

\author[0000-0001-7670-4897]{Ji-Hoon Ha}
\affil{Department of Physics, College of Natural Sciences UNIST, Ulsan 44919, Korea}
\author[0000-0002-5455-2957]{Dongsu Ryu}
\affil{Department of Physics, College of Natural Sciences UNIST, Ulsan 44919, Korea}
\author[0000-0002-4674-5687]{Hyesung Kang}
\affiliation{Department of Earth Sciences, Pusan National University, Busan 46241, Korea}
\author[0000-0002-5441-8985]{Sunjung Kim}
\affil{Department of Physics, College of Natural Sciences UNIST, Ulsan 44919, Korea}
\correspondingauthor{Ji-Hoon Ha, Dongsu Ryu}\email{hjhspace223@unist.ac.kr, dsryu@unist.ac.kr}

\begin{abstract}

Radio relics in the outskirts of galaxy clusters imply the diffusive shock acceleration (DSA) of electrons at merger-driven shocks with Mach number $M_{s}\lesssim3-4$ in the intracluster medium (ICM). 
Recent studies have suggested that electron preacceleration and injection, 
prerequisite steps for DSA, could occur at supercritical shocks with $M_{s}\gtrsim2.3$ in the ICM, thanks to the generation of multiscale waves by microinstabilities such as the Alfv\'en ion cyclotron (AIC) instability, the electron firehose instability (EFI), and the whistler instability (WI). 
On the other hand, some relics are observed to have subcritical shocks with $M_{s}\lesssim2.3$, leaving DSA at such weak shocks as an outstanding problem. 
Reacceleration of preexisting nonthermal electrons has been contemplated as one of possible solutions for that puzzle.
To explore this idea, we perform Particle-in-Cell (PIC) simulations for weak quasi-perpendicular shocks in high-$\beta$ ($\beta=P_{\rm gas}/P_{B}$) plasmas with power-law suprathermal electrons in addition to Maxwellian thermal electrons. 
We find that suprathermal electrons enhance the excitation of electron-scale waves via the EFI and WI. 
However, they do not affect the ion reflection and the ensuing generation of ion-scale waves via the AIC instability.
The presence of ion-scale waves is the key for the preacceleration of electrons up to the injection momentum, thus the shock criticality condition for electron injection to DSA is preserved.
Based on the results, we conclude that preexisting nonthermal electrons in the preshock region alone would not resolve the issue of electron preacceleration at subcritical ICM shocks.

\end{abstract}


\section{Introduction} 
\label{sec:s1}

Structure formation shocks are induced by supersonic flow motions during the hierarchical clustering of the large-scale structures of the universe including galaxy clusters  \citep[e.g.,][]{miniati2000, ryu2003, pfrommer2006, skillman2008,vazza2009,hong2014,schaal2015}. 
In particular, major mergers of clusters drive shocks with low sonic Mach numbers, $M_{s}\lesssim3-4$ \citep[e.g.,][]{ha2018a}, in the hot intracluster medium (ICM) of high plasma beta, $\beta\sim100$ \citep[e.g.,][]{ryu2008,porter2015}. Here, $\beta = P_{\rm gas}/P_{B}$, and $P_{\rm gas}$ and $P_{B}$ are the gas and magnetic pressures, respectively. Such merger-driven shocks are energetic enough to be detected in radio and X-ray observations \citep[e.g.,][]{markevitch2007, vanweeren2010, bruggen2012,brunetti2014}. 

In the so-called radio relics, the diffuse radio emission has been interpreted as synchrotron radiation from the cosmic-ray (CR) electrons produced by diffusive shock acceleration \citep[DSA; e.g.,][]{bell1978, blandford1978, drury1983} at merger-driven shocks \citep[see, e.g.,][for a review]{vanweeren2019}. Those shocks in the ICM are collisionless as in other astrophysical environments. With the shock obliquity angle between the shock normal and the background magnetic field, $\theta_{Bn}$, collisionless shocks are classified into ``quasi-perpendicular'' ($Q_{\perp}$, hereafter) if $\theta_{Bn} \gtrsim 45^{\circ}$ and ``quasi-parallel'' ($Q_{\parallel}$, hereafter) if $\theta_{Bn} \lesssim 45^{\circ}$. It is known from in-situ observations of Earth's bow shocks that electrons are energized mostly at $Q_{\perp}$ shocks \citep[e.g.,][]{gosling1989}. So the acceleration of CR electrons is expected to be effective primarily at the $Q_{\perp}$-population of merger shocks. On the other hand, using a high-resolution, long-integration, one-dimensional Particle-in-Cell (PIC) simulation, \citet{park2015} showed that, in ``strong'' $Q_{\parallel}$-shocks ($M_s\approx40$ or so), electrons could be accelerated via the Fermi acceleration mediated by ion-induced non-resonant Bell waves.

One of the key issues for electron acceleration at collisionless shocks is the so-called injection to DSA. This is because thermal electrons have the gyroradius much smaller than the shock thickness, and hence need to be ``preaccelerated'' to the ``injection momentum'', $p_{\rm inj}$, in order to diffuse across the shock transition and participate in the first-order Fermi process of DSA \citep[e.g.,][]{kang2002,capioli2015}. Here, the injection momentum is $p_{\rm inj} \gtrsim {\rm a~few} \times p_{i,{\rm th}}$, where $p_{i,{\rm th}} = \sqrt{2m_{i}k_{B}T_{i2}}$, $m_{i}$ is the ion mass, $k_{B}$ is the Boltzmann constant, and $T_{i2}$ is the downstream ion temperature. Hereafter, the subscripts 1 and 2 denote the quantities in the shock upstream and downstream regions, respectively. 

To study the electron preacceleration at ICM shocks, PIC simulations have been performed for $Q_{\perp}$-shocks with $M_s\lesssim$ several in plasmas of $\beta\approx20-100$ \citep[e.g.,][]{matsukiyo2011, matsukiyo2015, guo2014a, guo2014b, kang2019, niemiec2019, ha2021, kobzar2021}. 
In those simulations, two mechanisms involving wave-particle interactions have been identified for electron preacceleration: (1) multiple cycles of the shock drift acceleration (SDA) mediated by upstream self-excited waves, that is, the Fermi-like acceleration in the preshock region, and (2) the stochastic SDA (SSDA) mediated by waves of multiscales, ranging from electron to ion scales, in the shock transition zone. While the SDA of electrons, even after multiple cycles, would not be sufficient for electron injection, the SSDA could energize electrons
all the way up to $p_{\rm inj}$.

The plasma kinetic waves involved in these preacceleration processes are produced by a variety of microinstabilities. In collisionless $Q_{\perp}$-shocks, the instabilities are governed by the ions and electrons reflected or compressed at the shock ramp \citep[see, e.g.,][for more details]{kim2020, kim2021}. In the shock upstream, the reflected electrons induce the electron temperature anisotropy, $T_{e\parallel}/T_{e\perp}> 1$, which in turn excites the electron firehose instability (EFI). 
Scattering off the EFI-induced waves moderates a Fermi-like acceleration between the shock ramp and the upstream region \citep[e.g.][]{guo2014a,guo2014b,kang2019}.
In the shock transition zone, on the other hand, the ion temperature anisotropy is induced by the shock-reflected ions that are advected downstream, 
while the electron temperature anisotropy is induced by magnetic field compression at the shock ramp \citep[e.g.][]{guo2017,katou2019}. 
Then, those ion/electron temperature anisotropies could trigger instabilities of the following three types:
(1) the Alfv\'{e}n ion cyclotron (AIC) instability due to $T_{i\perp}/T_{i\parallel} > 1$, 
(2) the whistler instability (WI) due to $T_{e\perp}/T_{e\parallel} > 1$, 
and (3) the ion and electron mirror instabilities due to $T_{i\perp}/T_{i\parallel} > 1$ and $T_{e\perp}/T_{e\parallel} > 1$, respectively. 

The SSDA in the shock transition zone, which would be necessary for electron injection to DSA, relies on the extended confinement of electrons within the zone by scattering off multiscale waves and the resulting gradient-drift along the motional electric field,
and hence requires the effective excitation of the AIC instability \citep{niemiec2019,kobzar2021}. 
\citet[][Paper I, hereafter]{ha2021} showed that the reflection of incoming ions and the ensuing excitation of the AIC instability is expected to be effective only at  $Q_{\perp}$-shocks with $M_{s} \gtrsim M_{\rm crit}^{*} \approx 2.3$ in the ICM, 
implying that DSA would work only at those supercritical shocks.

The properties of merger-driven shocks in radio relics have been studied through observations. For instance, $M_s$ of the so-called ``radio relic shocks'' has been inferred either from the radio spectral index in radio observations \citep[e.g.,][]{vanweeren2010,vanweeren2019} or using the discontinuities in temperature or surface brightness in X-ray observations \citep[e.g.,][]{markevitch2002,markevitch2007}. 
In some radio relics, the Mach numbers inferred from radio observations is larger than those estimated from X-ray observations, i.e., $M_{\rm radio}\gtrsim M_{\rm X}$ \citep[e.g.,][]{akamatsu2013,vanweeren2019,wittor2021}.
It has been argued that the surfaces of merger shocks in the turbulent ICM are not uniform
and consist of parts with different $M_s$, and radio and X-ray observations preferentially pick up the parts of high and low Mach numbers, respectively \citep[e.g.,][]{hong2015,botteon2020,wittor2021}. 
This implies that there are intrinsic and inevitable uncertainties in $M_s$ estimated from observations of merger shocks.
Nevertheless, some radio relics seem to have shocks with $M_{s}$ close to or even smaller than $M_{\rm crit}^{*} \approx 2.3$, and examples include well-studied radio relics, such as those in ZwCL 0008.8+5215 \citep[e.g.,][]{kierdorf2017} and Abell 521 \citep[e.g.,][]{macario2013}. Hence, to explain the origin of radio relics, additional ingredients, other than the preacceleration and injection due to self-excited waves, would need to be considered.

As a possible solution for radio relics with low $M_s$ shocks, the so-called ``reacceleration model'' was suggested, where preexisting relativistic fossil electrons are injected to DSA and further accelerated \citep[e.g.,][]{kang2012,pinzke2013,kang2016a, kang2016b,kang2016c}. Those fossil electrons could enhance the acceleration efficiency, especially at weak shocks with $M_{s} \lesssim 3$ \citep[see, e.g.,][]{kang2011}. The presence of relativistic electron populations in the ICM is expected, possibly being supplied by radio jets of AGNs or produced through previous episodes of shock/turbulence acceleration \citep[e.g.,][and references therein]{kang2015, kang2016c}. As a matter of fact, some radio relics are observed to be associated with nearby radio galaxies \citep[e.g.,][]{vanweeren2017}, and low-frequency observations indicate the possibility of preexisting nonthermal electrons \citep[e.g.,][]{mandal2020}. 
However, the reacceleration scenario at weak shocks would operate, 
only if preexisting nonthermal electrons with $p\gtrsim p_{\rm inj}$
could be scattered diffusively across the shock transition.
It requires either the triggering of microinstabilities and
the self-generation of scattering waves or the presence of preexisting kinetic waves in the turbulent ICM.

In this paper, we revisit our previous work of PIC simulations for weak $Q_{\perp}$ ICM shocks, described in Paper I, by including power-law nonthermal electrons in addition to Maxwellian thermal electrons in the upstream plasma. Due to the limitation of computational capacities, the power-law component extends only up to the Lorenz factor of $\gamma\approx$ several, representing a suprathermal population, rather than the bonafide nonthermal population that continues to $\gamma\gg 1$. 
We study the effects of the preexisting suprathermal electrons on kinetic plasma processes in high-$\beta$ $Q_{\perp}$-shocks. 
In particular, we examine whether the suprathermal electrons could enhance the microinstabilities that generate multiscale waves, leading to the preacceleration of electrons up to $p_{\rm inj}$, even in subcritical shocks with $M_s\lesssim2.3$.

The paper is organized as follows. In the next section, simulations are described. The results are presented in Section \ref{sec:s3}. Summary and discussion follow in Section \ref{sec:s4}.

\begin{deluxetable*}{cccccccccccccc}[t]
\tablecaption{Model Parameters for Simulations \label{tab:t1}}
\tabletypesize{\scriptsize}
\tablecolumns{13}
\tablenum{1}
\tablewidth{0pt}
\tablehead{
\colhead{Model Name} &
\colhead{$M_{s}$} &
\colhead{$M_{A}$} &
\colhead{$u_0/c$} &
\colhead{$\theta_{Bn}$} &
\colhead{$\beta$} &
\colhead{$m_i/m_e$} &
\colhead{$\alpha$} &
\colhead{$n_{p}/n_{0}$}&
\colhead{$\gamma_{\rm cut}$}&
\colhead{$L_x[\lambda_{se}]$}&
\colhead{$L_y[\lambda_{se}]$}&
\colhead{$t_{\rm end} [\Omega_{ci}^{-1}]$}
}
\startdata
M2.0 &  2.0 & 8.2 & 0.038 & $63^{\circ}$ & 20 & 50 & 4.2 & 0.01 & 5 & 2000 & 200 & 22\\
M2.3 &  2.3 & 9.4 & 0.046& $63^{\circ}$  & 20 & 50 & 4.2& 0.01 & 5 & 2000 & 200& 22\\
M2.5 &  2.5 & 10.2 & 0.053& $63^{\circ}$ & 20 & 50 & 4.2& 0.01 & 5 & 2000 & 200 & 22\\
M3.0 & 3.0 & 12.3 & 0.068& $63^{\circ}$ & 20 & 50 & 4.2 & 0.01 & 5 & 2000 & 200 & 22\\
\hline
M2.0-$\alpha$4 &  2.0 & 8.2 & 0.038& $63^{\circ}$ & 20 & 50 & 4 & 0.01 & 5 &  2000 & 200  & 22\\
M2.0-$\alpha$4.5 &  2.0 & 8.2 & 0.038& $63^{\circ}$ & 20 & 50 & 4.5 & 0.01 & 5& 2000 & 200  & 22\\
M2.0-np0.1 & 2.0 & 8.2 & 0.038& $63^{\circ}$ & 20 & 50 & 4.2 & 0.1 & 5 & 2000 & 200  & 22\\
M2.0-$\gamma$cut10 &  2.0 & 8.2 & 0.038 & $63^{\circ}$ & 20 & 50 & 4.2 & 0.01 & 10 & 2000 & 200 & 22\\
M2.0-m25 &  2.0 & 8.2 & 0.053& $63^{\circ}$  & 20 & 25 & 4.2 & 0.01 & 5 & 2000 & 140  & 22\\
M2.0-$\beta$50-m25 &  2.0 & 12.9 & 0.053& $63^{\circ}$  & 50 & 25 & 4.2 & 0.01 & 5 & 2000 & 225  & 22\\
\hline
M3.0-$\alpha$4 &  3.0 & 12.3 & 0.068& $63^{\circ}$ & 20 & 50 & 4 & 0.01 & 5 &2000 & 200  & 22\\
M3.0-$\alpha$4.5 &  3.0 & 12.3 & 0.068& $63^{\circ}$ & 20 & 50 & 4.5 & 0.01 & 5 & 2000 & 200  & 22\\
M3.0-np0.1 & 3.0 & 12.3 & 0.068& $63^{\circ}$ & 20 & 50 & 4.2 & 0.1 & 5 & 2000 & 200  & 22\\
M3.0-$\gamma$cut10 & 3.0 & 12.3 & 0.068& $63^{\circ}$ & 20 & 50 & 4.2 & 0.01 & 10 & 2000 & 200 & 22\\
M3.0-m25 &  3.0 & 12.3 & 0.096& $63^{\circ}$ & 20 & 25 & 4.2 & 0.01 & 5 & 2000 & 140  & 22\\
M3.0-$\beta$50-m25 &  3.0 & 19.4 & 0.096& $63^{\circ}$ & 50 & 25 & 4.2 & 0.01 & 5 & 2000 & 225  & 22\\
\enddata
\vspace{-0.8cm}
\end{deluxetable*}

\section{Numerics}
\label{sec:s2}

Two-dimensional (2D) PIC simulations have been performed using TRISTAN - MP, a parallel electromagnetic PIC code \citep[][]{buneman1993,spitkovsky2005}. Here, we brief simulations, while further details can be found in Paper I. 
In the downstream rest frame, the background plasma composed with ions and electrons moves with the velocity $-u_0 \hat{x}$, and a collisionless shock propagating to the $+x$-direction is generated via interaction with reflected particles. The simulation domain is in the $x$-$y$ plane, and a uniform background magnetic field, ${\bf{B_0}}$, lies in the plane. We adopt the plasma parameters relevant for the ICM, the ion and electron temperatures $k_{B}T = k_{B}T_{i} = k_{B}T_{e} = 0.0168~m_ec^2 = 8.6$ keV (or $T_{i}=T_{e}=10^8$ K), the ion and electron number densities $n_0 = n_i = n_e=10^{-4} {\rm cm}^{-3}$, and the plasma beta $\beta = 16\pi n_0k_{B}T/B_0^2 = 20 - 50$. For given $\beta$ and $k_{B}T$, the sonic and Alfv\'{e}n Mach numbers are calculated as $M_{s} = u_{\rm sh}/\sqrt{2\Gamma k_{B}T_{i}/m_{i}}$ and $M_{A} = u_{\rm sh}/(B_0/\sqrt{4\pi n_0 m_{i}}) = M_{s}\sqrt{\Gamma \beta /2} $, respectively. Here, $\Gamma = 5/3$ is the adiabatic index. The shock velocity, $u_{\rm sh} = ru_0/(r - 1)$, is estimated using the shock compression ratio, $r = (\Gamma+1)/(\Gamma - 1 + 2/M_{s}^2)$. The sonic Mach number of simulated shocks is in the range of $M_{s} \sim 2 - 3$, covering subcritical and supercritical shocks, and $\theta_{Bn} = 63^{\circ}$ is chosen to emulate $Q_\perp$-shocks. 

The spatial resolution is $\Delta x = \Delta y = 0.1 \lambda_{se}$, and the time step is $\Delta t = 0.045 \omega_{pe}^{-1}$, where $\omega_{pe}=\sqrt{4\pi e^2n_0/m_e}$ and $\lambda_{se}=c/\omega_{pe}$ are the electron plasma frequency and the electron skin depth, respectively. Reduced mass ratios, $m_{i}/m_{e} = 25 - 50$, are employed to alleviate the demanding computational requirement. We note that $\Delta x$, $\Delta y$ and $\Delta t$ are expressed in units of the plasma skin depth, $\lambda_{se}$, and the electron plasma oscillation period, $\omega_{pe}^{-1}$. On the other hand, the shock changes on the time scale of the gyroperiod of incoming ions, $\Omega_{ci}^{-1} = m_i c/eB_0 \propto m_i \sqrt{\beta}$, and the shock transition develops on the length scale of their gyroradius, $r_{L,i} \equiv u_0 / \Omega_{ci} \approx 57\lambda_{se}\cdot (M_{s}/3)\sqrt{\beta/20}\sqrt{(m_i/m_e)/50}$. Hence, we use $\Omega_{ci}$ and $r_{L,i}$, when we describe the evolution and structure of the shock.

\begin{figure}[t]
\vskip 0.0 cm
\centerline{\includegraphics[width=0.45\textwidth]{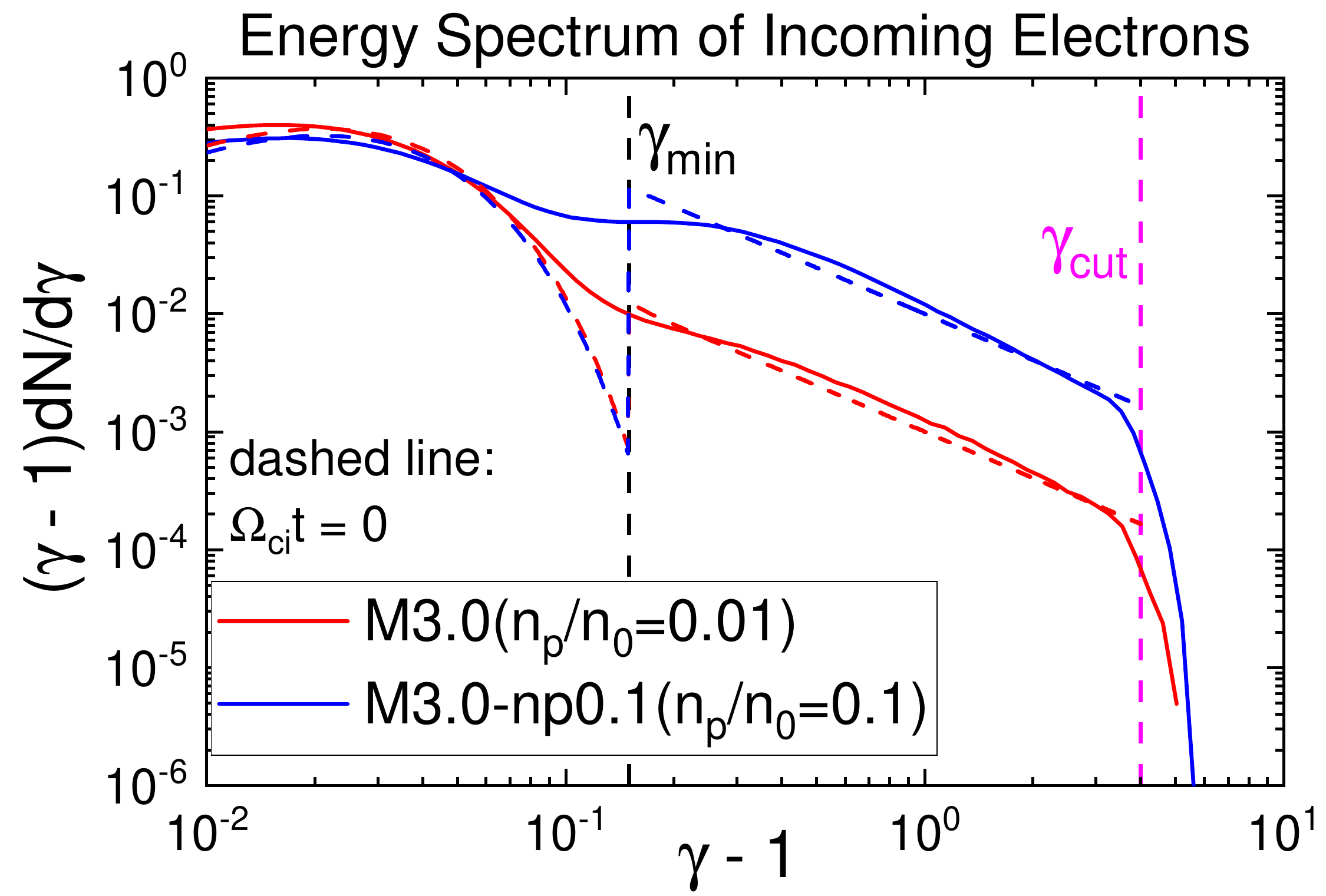}}
\vskip -0.2cm
\caption{Energy spectra of electrons in the far upstream region $10 \leq (x - x_{\rm sh})/r_{L,i} \leq 11$ (solid lines) and in the initial incoming plasmas (dashed lines) for two models M3.0 and M3.0-np0.1 (see Table \ref{tab:t1}). Here, $x_{\rm sh}$ is the shock position, and $r_{L,i} \approx 57\lambda_{se}$. The energy spectrum, $dN/d\gamma$, is related to the momentum spectrum, $f(p)\propto p^{-\alpha}$, as $4\pi f(p)p^2dp/dE \propto dN/d\gamma \propto (\gamma - 1)^{-s}$, where $s = \alpha - 2$.\label{fig:f1}}
\end{figure}

The main difference between this work and Paper I lies in the inclusion of 
preexisting power-law electrons (PPEs, hereafter). In this work, we add a nonthermal electron population of  ``isotropic'' power-law distribution, $f(p) \propto p^{-\alpha}$, to the background thermal plasma, reducing the amount of thermal electrons accordingly.
The power-law slope we consider covers the range of $\alpha = 4 - 4.5$, 
and the number fraction of PPEs ranges $n_{p}/n_{0} = 0.01 - 0.1$.
For the fiducial model, $\alpha = 4.2$ is adopted, assuming that the preexisting nonthermal electrons are produced by relativistic shocks such as those in radio jets \citep[e.g.,][]{kirk2000, achterberg2001}. 
The minimum and maximum (or cutoff) Lorenz factors of the nonthermal component are $\gamma_{\rm min} = 1.15$ (corresponding to the velocity $\sim 0.5c$) and $\gamma_{\rm cut} = 5 - 10$, respectively. 
Figure \ref{fig:f1} shows the energy spectra of electrons measured in the far upstream region of $10 \leq (x - x_{\rm sh})/r_{L,i} \leq 11$ at $\Omega_{ci}t \sim 3$, and compares them with those of the initial incoming plasmas.
Here, $x_{\rm sh}$ denotes the position of the shock. We see a smoothing in the upstream energy spectra, resulting in the continuous merging of the power-law component to the Maxwellian component. Although it is not further discussed here, the whistler mode is responsible for it. Even with the smoothing, the power-law distribution is pretty well maintained.

In PIC simulations, the limited number of particles induces numerical dissipations \citep[e.g.,][]{melzani2013}, and hence a sufficiently large number of particles is required to maintain the power-law distribution of electrons for a sufficiently long time. In this work, 120 particles per cell (60 per species) are placed, which are much larger than 32 particles per cell (16 per species) used in Paper I. Even with this number of particles, $\gamma_{\rm cut} = 5 - 10$ is the maximum value that we can accommodate. 
We point that this value of $\gamma_{\rm cut}$ is set by the computational limitation, rather than the physical argument. 
With $\gamma_{\rm cut} = 5 - 10$, the PPE component should be regarded in effect as a suprathermal population. 

The model parameters of our simulations are given in Table \ref{tab:t1}. Models with different $M_{s}$ are named with the combination of the letter `M' and the sonic Mach number (for instance, the M3.0 model has $M_{s}=3$). The four models in the top group are the fiducial models, which have $\beta=20$, $m_i/m_e=50$, $\alpha = 4.2$, $n_{p}/n_0 = 0.01$, and $\gamma_{\rm cut} = 5$. Models with parameters different from those of the fiducial models have the names that are appended by a character for the specific parameter and its value; for example, the M3.0-np0.1 model has $n_{p}/n_0 = 0.1$, while the M3.0-$\alpha4$ has the power-law slope $\alpha = 4$. For the box size, $L_x$ and $L_y$, the end time of simulations, $t_{\rm end}$, and $\beta$, we adopt the values somewhat smaller than those of Paper I to compensate the longer computational time due to the larger number of particles. 
Yet, the adopted values of these parameters should be pertinent enough to capture the main results of this work. 
In the next section, the fiducial models in the current work will be compared with the models with $\beta = 20$ in Paper I (without PPEs).

\begin{figure*}[t]
\vskip 0.0 cm
\centerline{\includegraphics[width=1.0\textwidth]{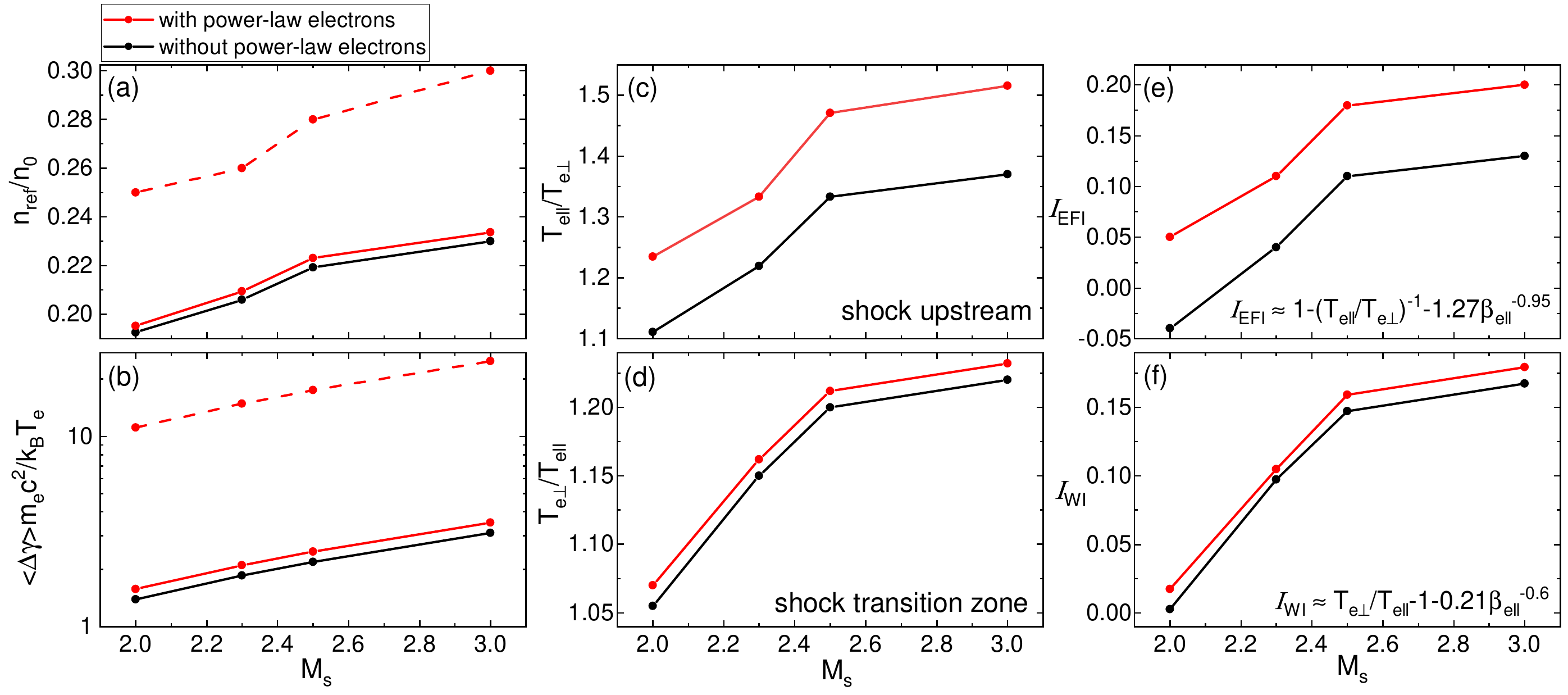}}
\vskip -0.2cm
\caption{Fraction of reflected electrons (panel (a)) and average energy gained by reflected electrons in a single cycle of SDA (panel (b)), based on the relativistic SDA description \citep{guo2014a,kang2019}. Electron temperature anisotropy, $T_{e\parallel}/T_{e\perp}$, estimated in the immediate upstream region, $0 \leq (x - x_{\rm sh})/r_{L,i} \leq 1$ (panel (c)), and $T_{e\perp}/T_{e\parallel}$ estimated in the immediate downstream region, $-1 \leq (x - x_{\rm sh})/r_{L,i} \leq 0$ (panel (d)). Instability parameters for the EFI, $I_{\rm EFI}$ (panel (e)), and for the WI, $I_{\rm WI}$ (panel (f)). The red solid lines draw the fiducial models with PPEs, while the black solid lines draw the corresponding models without PPEs from Paper I, at $\Omega_{ci}t \sim 22$. The red dashed lines in panels (a) and (b) show the quantities estimated only with PPEs, that is, the fraction of reflected PPEs and the average energy gained by reflected PPEs in a single cycle of SDA. \label{fig:f2}}
\end{figure*}

\begin{figure*}[t]
\vskip 0.0 cm
\centerline{\includegraphics[width=1.0\textwidth]{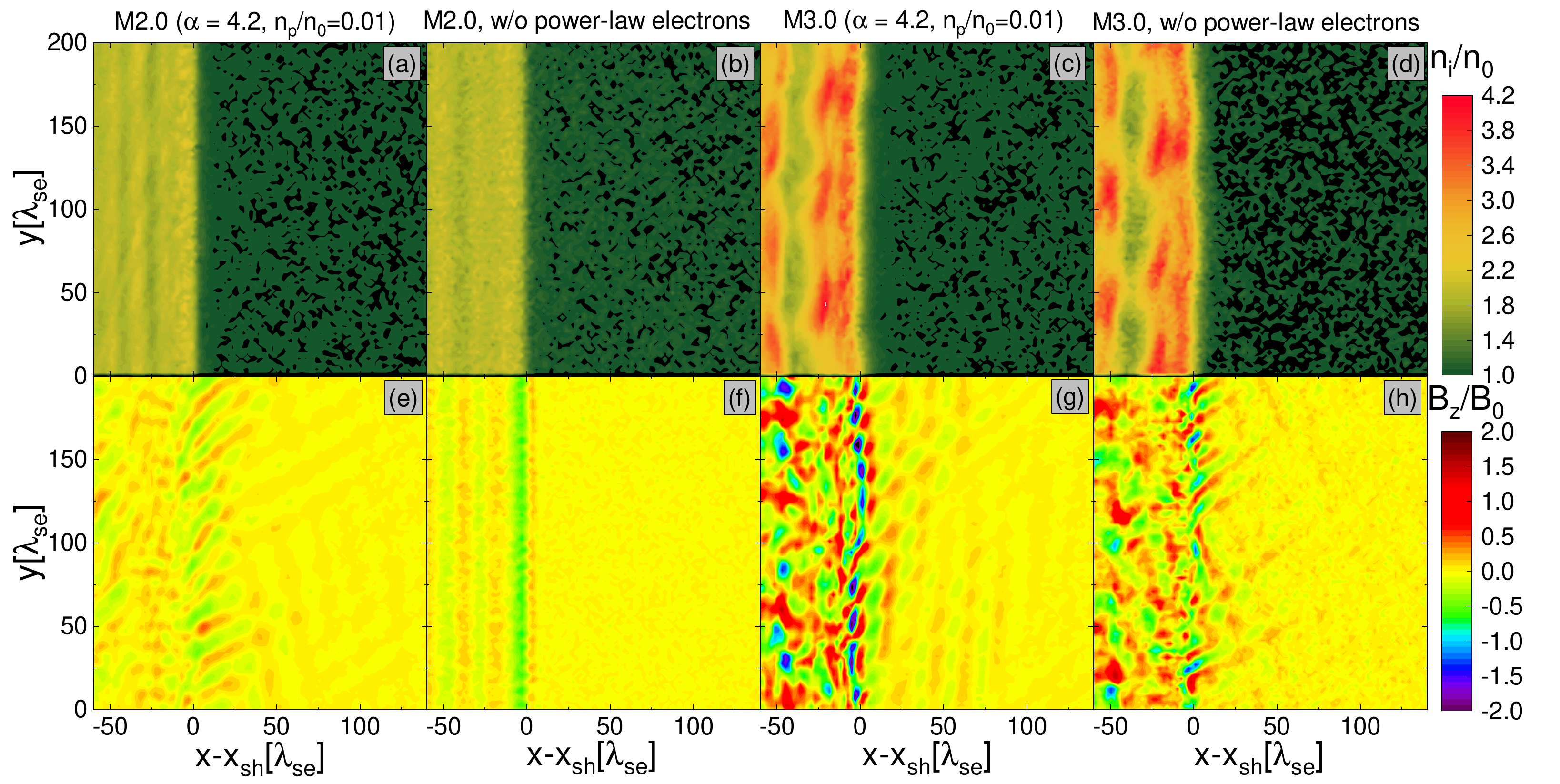}}
\vskip -0.2cm
\caption{Ion number density (top panels), $n_{i}$, normalized to the upstream ion number density, $n_0$, and $z$-magnetic field (bottom panels), $B_{z}$, normalized to the upstream magnetic field, $B_0$, in the region of $-60 \leq (x - x_{\rm sh})/\lambda_{se} \leq 140$ around the shock at $\Omega_{ci}t \sim 22$. Here, $x_{\rm sh}$ is the shock position. The fiducial M2.0 and M3.0 models with $\alpha = 4.2$ and $n_{p}/n_0 = 0.01$ (panels (a), (e), (c), (g)) are compared to the corresponding models without PPEs from Paper I (panels (b), (f), (d), (h)). \label{fig:f3}}
\end{figure*}

\begin{figure*}[t]
\vskip 0.0 cm
\hskip -0.2 cm
\centerline{\includegraphics[width=0.75\textwidth]{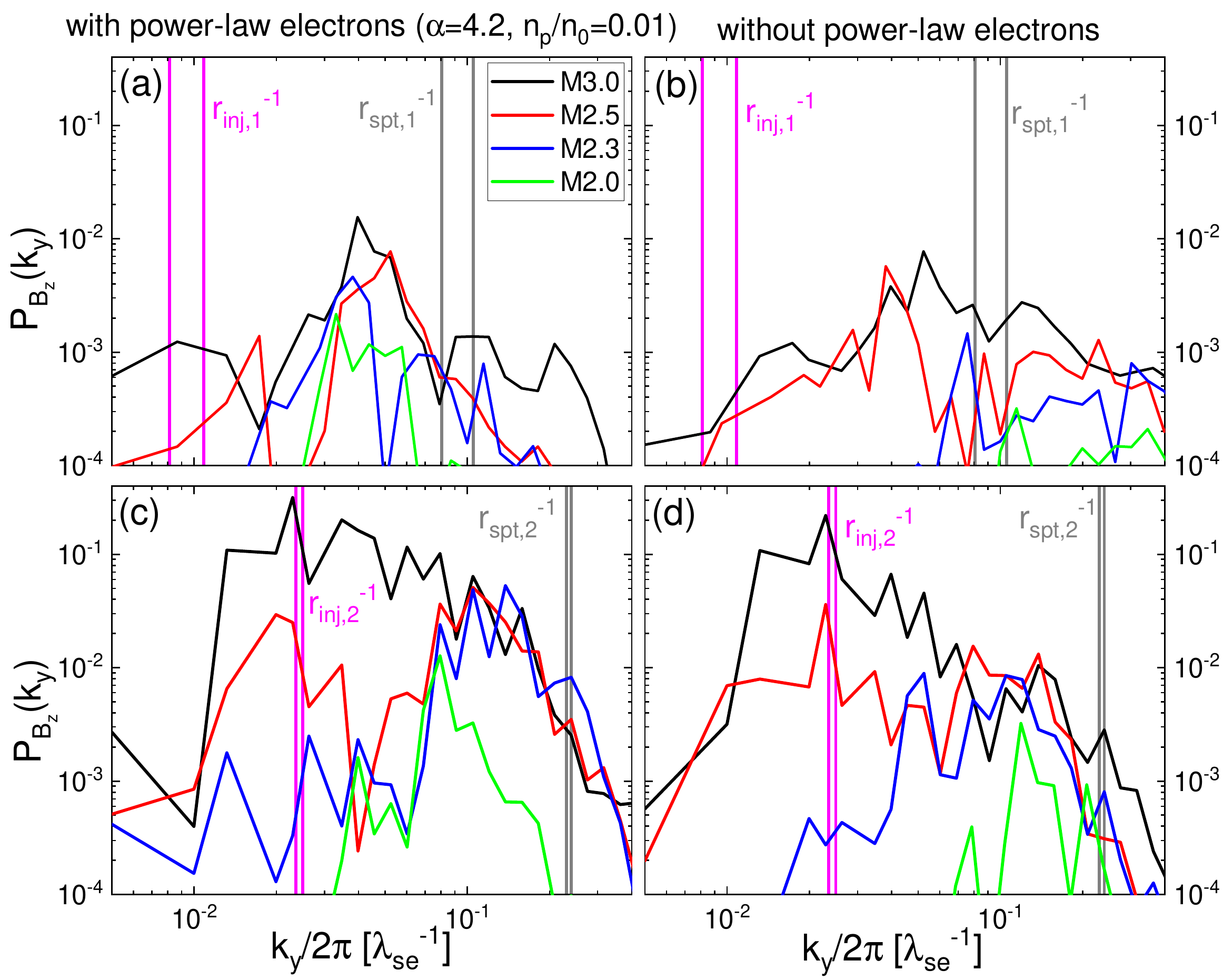}}
\vskip -0.2cm
\caption{Magnetic field power spectra, $P_{B_z}(k_{y}) \propto (k_y/2\pi)(\delta B_z(k_y)^2/B_0^2)$, in the immediate upstream region, $0 \leq (x - x_{\rm sh})/r_{L,i} \leq 1$ (top panels), and the immediate downstream region, $-1 \leq (x - x_{\rm sh})/r_{L,i} \leq 0$ (bottom panels), at $\Omega_{ci}t \sim 22$, for the models with PPEs (left panels) and without PPEs (right panels).
Here, $x_{\rm sh}$ is the shock position, and $r_{L,i} \approx 57\lambda_{se}\cdot (M_{s}/3)$. 
The magenta and gray boxes denote the ranges of the gyroradius of electrons with $p_{\rm inj}$ and $p_{\rm spt} \equiv \sqrt{m_e/m_i}~p_{\rm inj}$, respectively, in the upstream and downstream regions.\label{fig:f4}}
\end{figure*}

\section{Results}
\label{sec:s3}

\subsection{Particle Reflection and Temperature Anisotropies}
\label{s3.1}

According to the description of relativistic SDA theory for oblique shocks \citep[e.g.,][]{guo2014a,kang2019}, the amount of reflected electrons is expected to increase, if PPEs with higher velocities are added to the Maxwellian distribution. We first examine its consequences and effects on SDA and instabilities in Figure \ref{fig:f2}. In Figure \ref{fig:f2}(a,b), the fraction of reflected electrons and the average energy gained by those reflected electrons in a single SDA cycle are shown. Indeed, the fraction of reflected PPEs (red dashed line) is larger than that of reflected thermal electrons (black solid line), so is the fraction of total reflected electrons in the models with PPEs (red solid line). The average energy gained by reflected PPEs in a single SDA cycle is larger than that of thermal electrons as well, because electrons with larger $\gamma$ drift longer distances along the shock surface. Hence, with PPEs, the electron temperature anisotropies, $T_{e\parallel}/T_{e\perp}$, estimated in the shock upstream, $0 \leq (x - x_{\rm sh})/r_{L,i} \leq 1$, increase (Figure \ref{fig:f2}(c)), and then based on the instability condition for the EFI, $I_{\rm EFI} \approx 1 - (T_{\rm e\parallel}/T_{\rm e\perp})^{-1} - {1.27}{\beta_{\rm e\parallel}^{-0.95}}>0$ \citep[see, e.g.,][]{kang2019}, the growth of the EFI is expected to be enhanced in the upstream region (Figure \ref{fig:f2}(e)). On the other hand, the SDA-accelerated electrons and PPEs, which are advected downstream, would lead to a slight increase of the electron temperature anisotropy, $T_{e\perp}/T_{e\parallel}$, in the shock transition zone, $-1 \leq (x - x_{\rm sh})/r_{L,i} \leq 0$, (Figure \ref{fig:f2}(d)). Then, according to the instability condition for the WI, $I_{\rm WI} \approx T_{\rm e\perp}/T_{\rm e\parallel} - 1 - {0.21}{\beta_{\rm e\parallel}^{-0.6}}>0$ \citep[see, e.g.,][]{guo2017}, the growth of the WI would be slightly enhanced in the transition zone (Figure \ref{fig:f2}(f)).

The instability condition for the AIC,
$I_{\rm AIC} \approx (T_{\rm i \perp} /T_{\rm i \parallel})-1 - {1.6}{\beta_{\rm i\parallel}^{-0.72}}>0$ (Paper I),
depends on the temperature anisotropy of ions, which is induced by the gyrating reflected ions in
the plane perpendicular to the shock-compressed magnetic field \citep{guo2017}.
Obviously, the presence of PPEs would not affect the ion reflection, 
so we expect that it would not increase the ion temperature anisotropy, nor enhance the growth of the AIC
in the shock transition zone.

\subsection{Generation of Waves by Microinstabilities}
\label{s3.2}

Figure \ref{fig:f3} shows the distributions of the ion number density, $n_i$ (top panels), and the self-excited magnetic field, $B_z$ (bottom panels), around the shock for the M2.0 and M3.0 models with and without PPEs. 
Oblique waves with $\lambda \sim 20 - 30 \lambda_{se}$ are generated in the preshock region of the M2.0 model with PPEs (Figure \ref{fig:f3}(e)), while such waves are absent in the M2.0 model without PPEs (Figure \ref{fig:f3}(f)). 
They are expected to be induced by the EFI, and present also in the upstream region of the M3.0 models both with and without PPEs (Figure \ref{fig:f3}(g, h)).
\citet{kang2019} demonstrated that the fraction of reflected, backstreaming electrons, the energy gain via SDA, and the temperature anisotropy, $T_{e\parallel}/T_{e\perp}$, are large enough to trigger the EFI only in supercritical shocks with $M_s\gtrsim2.3$. 
If PPEs are added to the incoming plasma, on the other hand, the fraction of reflected electrons and $T_{e\parallel}/T_{e\perp}$ increase as shown in Figure \ref{fig:f2}(a,c), and thus the EFI is excited even in the subcritical shock with $M_s=2$ (Figure \ref{fig:f3}(e)).
Moreover, the comparison of the models with and without PPEs reveals that electron-scale waves along the overshoot, excited by the WI, are somewhat enhanced by PPEs as well.

In Paper I, it was shown that in supercritical shocks, the AIC instability due to the ion temperature anisotropy could excite
ion-scale waves with $\lambda_{\rm ripple} \sim 70 \lambda_{se}$, leading to ripples propagating along
the shock surface (Figure \ref{fig:f3} (c,d)).
As a result, multiscale plasma waves, ranging from electron to ion scales, appear in the supercritical M3.0 models (Figure \ref{fig:f3} (g,h)). 
On the contrary, the emergence of such ripples and ion-scale waves 
are not observed in the subcritical M2.0 models, regardless of PPEs (Figure \ref{fig:f3} (a,b)).

The presence of plasma waves on relevant kinetic scales is essential for electron preacceleration and subsequent injection to DSA. This is because
for electron energization up to $p_{\rm inj}$, wave-particle interactions mediated by waves with wavelengths as long as the gyroradius of electrons with $p_{\rm inj}$ are required. The injection momentum ranges $p_{\rm inj} \approx 3 \sqrt{2m_{i}k_{B}T_{2}} \approx 5.6 - 7.5 m_{e}c$ 
for the fiducial models with $M_{s} = 2 - 3$, $T_{1} = 10^8$ K, and $m_{i}/m_{e} = 50$. 
Hence, the gyroradius of electrons with $\gamma_{\rm inj}$ is $r_{\rm inj,1} \equiv \gamma_{\rm inj}c/\Omega_{ce,1} \approx 92 - 123 \lambda_{se}$ and $r_{\rm inj,2} \equiv \gamma_{\rm inj}c/\Omega_{ce,2} \approx 40 - 41 \lambda_{se}$ in the preshock and postshock regions, respectively. Here, $\Omega_{ce,1}$ and $\Omega_{ce,2}$ are the gyrofrequencies of electrons. 
Figure \ref{fig:f3} demonstrates that in subcritical shocks, such ion-scale waves with $\lambda \sim r_{\rm inj,1}$ or $r_{\rm inj,2}$ are not excited by the addition of PPEs (Figure \ref{fig:f3}(a,e)).

\begin{figure}[t]
\vskip 0.0 cm
\centerline{\includegraphics[width=0.43\textwidth]{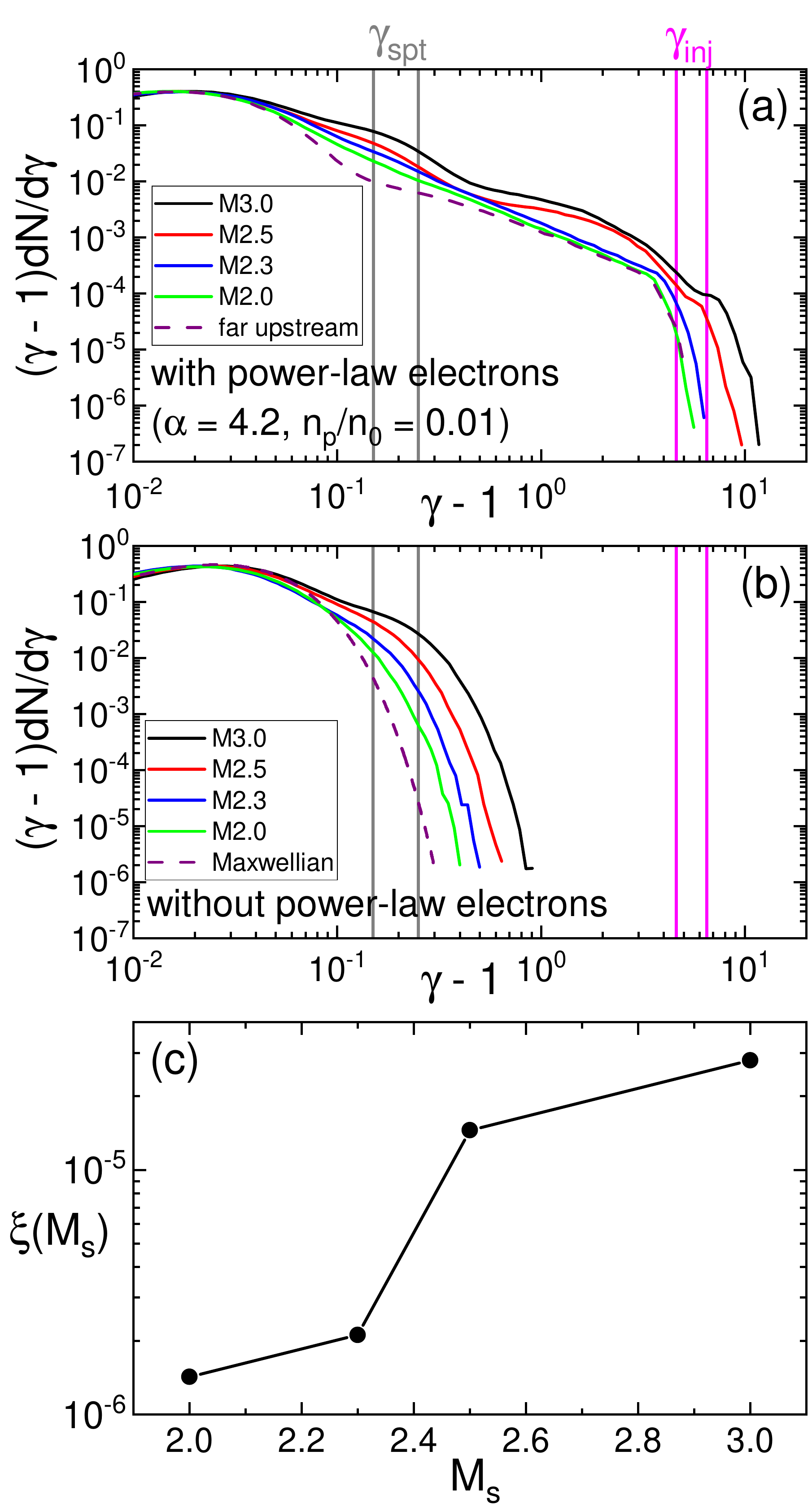}}
\vskip -0.3 cm
\caption{Electron energy spectra in the immediate upstream region, $0 \leq (x - x_{\rm sh})/r_{L,i} \leq 1$, at $\Omega_{ci}t \sim 22$, for the models with PPEs (panel (a)) and without PPEs (panel (b)). The dark purple dashed lines draw either the far upstream spectrum shown in Figure \ref{fig:f1} (panel (a)) or the Maxwellian spectrum (panel (b)). 
The magenta and gray boxes denote the ranges of $\gamma_{\rm inj}$ and $\gamma_{\rm spt}$, respectively. 
Panel (c) shows the injection fraction of electrons defined in Equation (\ref{eq1}), estimated for the fiducial models with PPEs in panel (a). Here, $x_{\rm sh}$ is the shock position, and $r_{L,i} \approx 57\lambda_{se} \cdot (M_{s}/3)$. \label{fig:f5}}
\end{figure}

For the further characterization of waves, the power spectra of $B_z$, $P_{B_{z}}$, in the immediate upstream (top panels) and downstream (bottom panels) regions of the shock are shown in Figure \ref{fig:f4}, for all the fiducial models with PPEs in Table \ref{tab:t1} (left panels) and the corresponding models without PPEs (right panels). The ranges of $r_{\rm inj}$ (magenta) and the gyroradius, $r_{\rm spt}$, for the suprathermal momentum, $p_{\rm spt}\equiv \sqrt{m_e/m_i}~p_{\rm inj}$, (gray) are plotted for reference.
Note that $p_{\rm spt}$ roughly marks the momentum above which electrons change from thermal to suprathermal distributions \citep{kang2019}. 
The comparison of the subcritical shocks (green and blue lines) with and without PPEs
shows that PPEs enhance waves mainly with $\lambda \sim 20-30 \lambda_{se} \ll r_{\rm inj,1},r_{\rm inj,2} $,
which cannot mediate the energization of electrons up to $p_{\rm inj}$. 
The downstream waves with $r_{\rm spt,2}<\lambda<r_{\rm inj,2}$ appear partly due to the WI at the shock transition and partly via the advection of upstream waves.

On the other hand, waves with $\lambda\gtrsim r_{\rm inj,2}$ are produced only in the supercritical shocks (black and red lines), 
regardless of the presence of PPEs.
An interesting point is that in the marginally critical shock of $M_s=2.3$, waves with $\lambda\gtrsim r_{\rm inj,2}$ are enhanced slightly with PPEs, possibly owing to the modification of magnetic field fluctuations by the EFI mode. 
Again, Figure \ref{fig:f4} confirms that the multiscale plasma waves that can resonate with electrons of $p_{\rm spt}\lesssim p \lesssim  p_{\rm inj}$
are produced only at supercritical shocks with $M_s\gtrsim2.3$.

\subsection{Energy Spectrum and Injection Fraction}
\label{s3.3}

The consequences of wave-particle interactions should be manifested in the electron energy spectra. 
Figure \ref{fig:f5} shows the spectra measured in the shock upstream for all the models with PPEs (Figure \ref{fig:f5} (a)) and without PPEs (Figure \ref{fig:f5} (b)).
The spectra in Figure \ref{fig:f5}(b) are basically the same as those in Paper I.
In the M2.0 model without PPEs (green line in Figure \ref{fig:f5}(b)),
the EFI is not triggered, and so only a single SDA cycle occurs.
In the M3.0 model without PPEs (black line in Figure \ref{fig:f5}(b)), on the other hand, 
thermal electrons are energized up to $\gamma\sim2$ via the Fermi-like acceleration and SSDA.

In the spectra for the models with PPEs in Figure \ref{fig:f5}(a), two points are noted. 
(1) The suprathermal population grows in all the models, even in the M2.0 model. 
Especially, the population with $\gamma \lesssim 1.5$ represents the incoming thermal electrons energized via the Fermi-like acceleration mediated by EFI-driven waves. 
(2) In the supercritical models, the spectra stretch beyond $\gamma_{\rm inj}$ through the SSDA mediated by multiscale waves. 
However, in the M2.0 model, the high-energy end of the spectrum does not change, because PPEs do not aid the excitation of AIC-driven ion-scale waves.
These results demonstrate that the presence of PPEs alone does not modify the critical Mach number, $M_{\rm crit}\approx 2.3$, for electron preacceleration and subsequent injection to DSA.

In PIC simulations, the maximum energy of electrons is limited by computational constraints; hence, with a larger simulation box and a longer integration time, the spectra would extend to higher energies, as noted in Paper I. 
Previously, \citet{trotta2019} performed hybrid simulations, implemented with test-particle electrons of initially kappa distributions. In their work, electron-scale waves, induced by the EFI and whistler modes, are not present due to the lack of electron dynamics.
Yet, they showed that in supercritical shocks, electrons could be energized well above $p_{\rm inj}$ through interactions with multiscale waves, which are accompanied by the shock surface rippling triggered by the ion dynamics. 

To quantify electron preacceleration, we estimate the injection fraction, $\xi$, defined as
\begin{equation}
\xi \equiv \int_{\gamma_{\rm inj}}^{\infty}\frac{dN}{d\gamma}d\gamma ~ \Bigg/ \int_1^{\infty}\frac{dN}{d\gamma}d\gamma,
\label{eq1}
\end{equation}
in the shock upstream, $0 \leq (x - x_{\rm sh})/r_{L,i} \leq 1$.
Figure \ref{fig:f5}(c) shows $\xi(M_s)$ for the fiducial models in Table \ref{tab:t1}.
As expected, $\xi$ increases sharply at $M_s\gtrsim2.3$, marking the critical Mach number. 
We point that $\xi$ shown in Figure \ref{fig:f5}(c) is not necessarily the injection fraction of electrons at $Q_{\perp}$-shocks in the ICM.
Our PIC simulations are limited only to reproduce processes on kinetic scales. 
With a much larger spatial domain and a much longer dynamical time, both thermal and suprathermal electrons are expected to be preaccelerated and injected to DSA in supercritical shocks, hence the injection fraction would be substantially larger than $\xi$ in Figure \ref{fig:f5}(c). 
While the estimation of realistic injection fractions is beyond the scope of this paper, we conjecture that $\xi$ should be small at subcritical shocks unless there are additional processes to boost it. 

\begin{figure}[t]
\vskip 0.0 cm
\centerline{\includegraphics[width=0.43\textwidth]{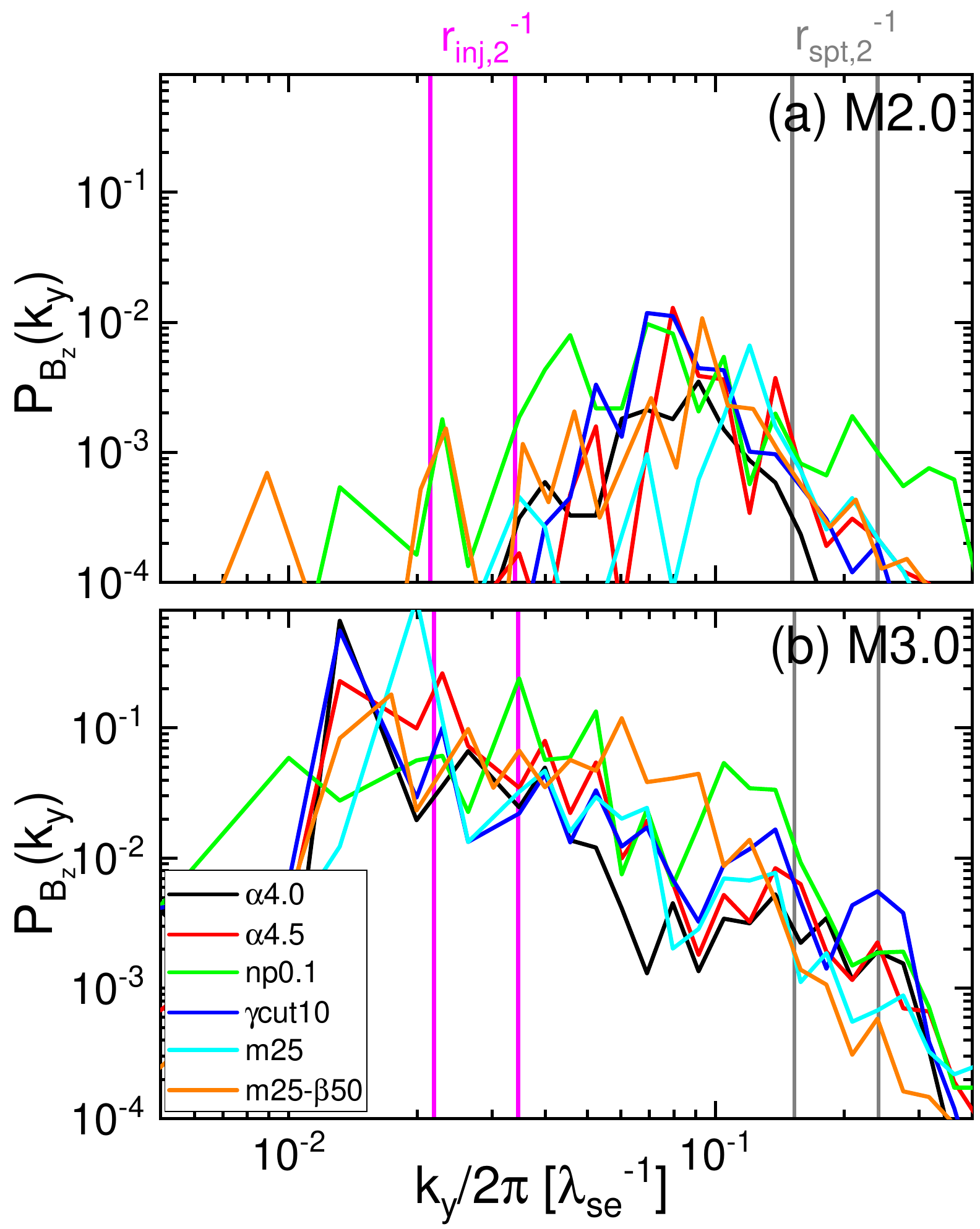}}
\vskip -0.1 cm
\caption{Magnetic field power spectra, $P_{B_z}(k_{y}) \propto (k_y/2\pi)(\delta B_z(k_y)^2/B_0^2)$, in the immediate downstream region, $-1 \leq (x - x_{\rm sh})/r_{L,i} \leq 0$, at $\Omega_{ci}t \sim 22$, for the M2.0 (top) and M3.0 (bottom) models with different parameters (see Table \ref{tab:t1}).
Here, $x_{\rm sh}$ is the shock position, and $r_{L,i} \approx 57\lambda_{se} \cdot (M_{s}/3)\sqrt{\beta/20}\sqrt{(m_i/m_e)/50}$.
The magenta and gray boxes denote the ranges of the gyroradius of electrons with $p_{\rm inj}$ and $p_{\rm spt}$, respectively, in the downstream region. \label{fig:f6}}
\end{figure}

\begin{figure*}[t]
\vskip 0.0 cm
\centerline{\includegraphics[width=0.9\textwidth]{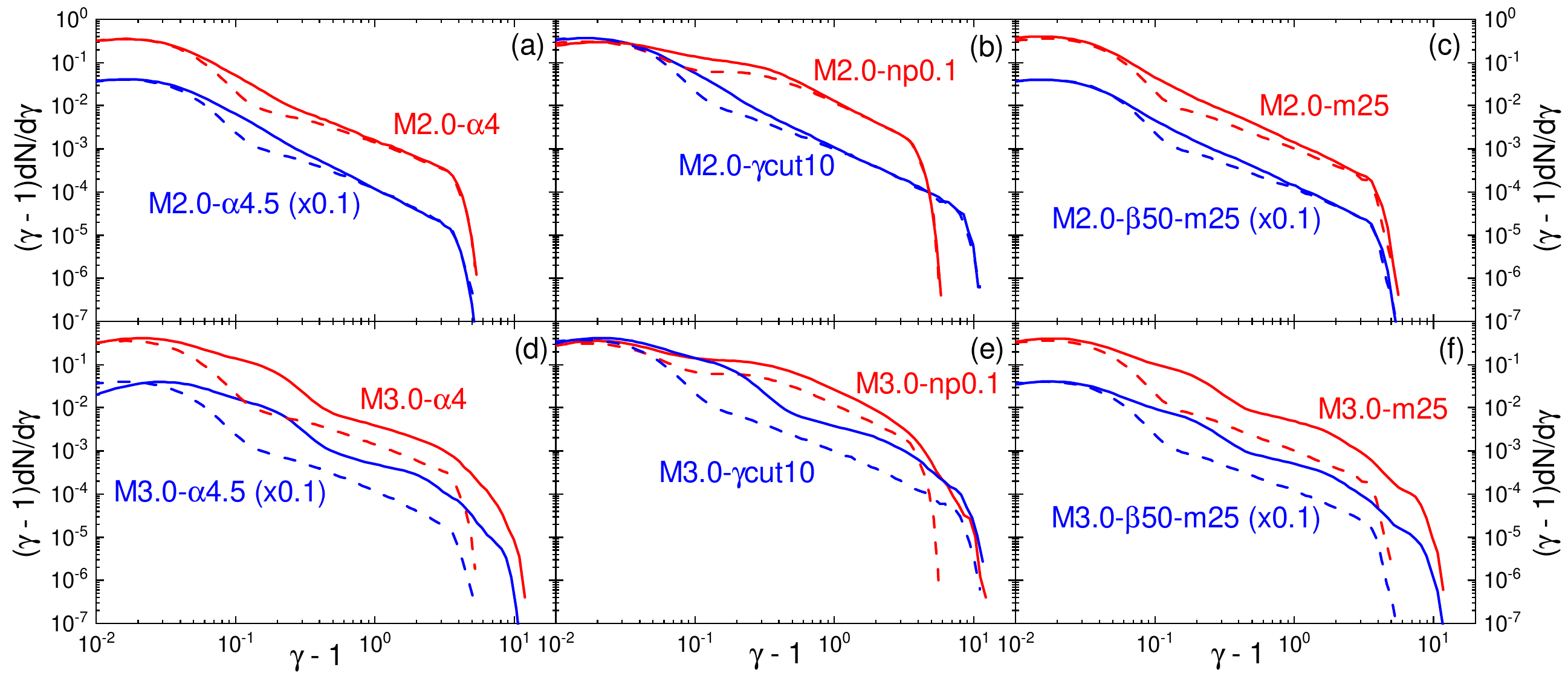}}
\vskip -0.2cm
\caption{Electron energy spectra in the immediate upstream region, $0 \leq (x - x_{\rm sh})/r_{L,i} \leq 1$, at $\Omega_{ci}t \sim 22$, for the M2.0 (upper panels) and M3.0 (lower panels) models with different parameters (see Table \ref{tab:t1}).
The dashed lines draw the electron energy spectra in the far upstream region, $10 \leq (x - x_{\rm sh})/r_{L,i} \leq 11$, at $\Omega_{ci}t \sim 3$, for comparison.
Here, $x_{\rm sh}$ is the shock position, and $r_{L,i} \approx 57\lambda_{se} \cdot (M_{s}/3)\sqrt{\beta/20}\sqrt{(m_i/m_e)/50}$. 
Some of the blue solid and dashed lines marked by ($\times 0.1$) are shifted vertically for clarity.
\label{fig:f7}}
\end{figure*}

\subsection{Dependence on Parameters} 
\label{s3.4}

We next examine how our findings depend on the parameters that specify PPEs, such as $n_{p}/n_{0}$, $\alpha$, and $\gamma_{\rm max}$, as well as the simulation parameters, such as $m_i/m_e$ and $\beta$. 
Figure \ref{fig:f6} shows the magnetic field power spectra, $P_{B_{z}}$, in the immediate downstream, for the M2.0 models in the second group (Figure \ref{fig:f6}(a)) and the M3.0 models in the third group (Figure \ref{fig:f6}(b)) of Table \ref{tab:t1}.  
The comparison of all the lines in Figure \ref{fig:f6}(b) with the black line in Figure \ref{fig:f4}(c) indicates
that the generation of waves does not strongly depend on these parameters in the case of the supercritical shocks with $M_s=3$. 
In all the M3.0 models, $P_{B_{z}}$ for $r_{\rm spt,2}\lesssim \lambda \lesssim r_{\rm inj,2}$ is
substantial. By contrast, there are some differences in the case of the subcritical M2.0 models.
For instance, in the M2.0-np0.1 model (green line in Figure \ref{fig:f6}(a)), with $n_p/n_0$ ten times larger than that of the fiducial model (green line in Figure \ref{fig:f4}(a)), the excitation of waves is enhanced. 
In the M2.0-$\beta$50-m25 model (orange line in Figure \ref{fig:f6}(a)), with a weaker background magnetic field, waves with longer wavelengths appear.
As noted above, waves in the M2.0 models are mostly electron-scale waves, induced by the EFI and whistler modes. 

Finally, Figure \ref{fig:f7} compares the electron energy spectra in the immediate upstream for the M2.0 models in the second group (upper panels) and the M3.0 models in the third group (lower panels) of Table \ref{tab:t1}. 
The energy spectra extend up to $\gamma-1\sim10$ in all the M3.0 models, independent of the parameters.
This is because the ripples along the shock surface have $\lambda_{\rm ripple} \sim 70 \lambda_{se}$ (see Figure \ref{fig:f3}(c,d)), which marks the longest wavelength of AIC-driven waves. 
Hence, the maximum energy in the spectra is given as $\gamma_{\rm max} \approx \lambda_{\rm ripple}\Omega_{ce,2}/c \approx 11$, in our simulations. 
Again, the main results described in the previous subsection remain valid, independent of the model parameters.

\section{Summary and Discussion}
\label{sec:s4}

Electrons are expected to be preaccelerated and injected to the Fermi-I, DSA process in supercritical $Q_{\perp}$-shocks with $M_s \gtrsim 2.3$ in the high-$\beta$ ICM \citep[][and Paper I]{kang2019}.
Here, we explore if the presence of preexisting nonthermal electrons could facilitate the generation of kinetic waves
via microinstabilities and assist the electron injection to DSA, especially in the subcritical regime.
We have performed 2D PIC simulations for the following parameters: $M_s=2-3$, $\theta_{\rm Bn}=63^{\circ}$,
and $\beta=20$ (see Table \ref{tab:t1}).
Preexisting nonthermal electrons are represented by a power-law component (PPEs) that extends up to $\gamma_{\rm cut}=5-10$.

The main findings are summarized as follows. (1) The presence of PPEs enhances the EFI owing to the increased fraction of the electrons reflected at the shock ramp. 
As a result, waves with $\lambda\sim20-30\lambda_{se}$, generated by the EFI, appear even in the upstream region of the subcritical shocks in our M2.0 models. 
At the same time, the suprathermal population increases through the preacceleration via 
the Fermi-like acceleration mediated by the EFI waves in the M2.0 models, as well as in other models with supercritical shocks. 
(2) Ion-scale waves with $\lambda\gtrsim r_{L,i} \approx 57\lambda_{se}\cdot (M_{s}/3)$ are excited only in the supercritical shocks of the M2.3 - M3.0 models, as the consequence of the AIC instability, regardless of the presence of PPEs.
In those models, the energy spectrum of electrons continues beyond $p_{\rm inj}$ via the SSDA mediated by multiscale waves ranging from electron to ion scales. 
On the other hand, ion-scale waves are not produced in the subcritical M2.0 models,
since PPEs do not facilitate the excitation of ion-scale waves via the AIC instability.
Hence, in the M2.0 models, the high-energy end of the electron energy spectrum does not change even with PPEs.

We further comment on the potential effects of the three ingredients that are not fully accounted for in our simulations: (1) realistic preexisting nonthermal electrons, (2) preexisting nonthermal ions, 
and (3) preexisting kinetic turbulence in the ICM.

As noted above, while the preexisting power-law component of electrons extends only up to $\gamma\lesssim\gamma_{\rm inj}$ in our simulations due to limited computational capacities, the nonthermal population in the ICM is expected to stretch to higher energies of $\gamma\sim10^2-10^3$, considering the physical condition there \citep[e.g.,][]{sarazin1999,mandal2020}. Hence, in reality, some of the electrons reflected at the shock ramp may have energies higher than those reproduced in our simulations. On the other hand, the properties of the EFI induced by the reflected electrons, such as the wavelength and growth rate of the fastest growing mode, are not very sensitive to the characteristics of the electron beam that induces the instability \citep[see, e.g.,][]{kim2020}; as long as the bulk of the beam energy resides in the low-energy part of the power-law distribution, most of the EFI-induced waves would be still on electron scales. Although the details should be further investigated, we suspect that the power-law nonthermal electrons that extend to higher energies would not resolve the issue of the generation of ion-scale waves in subcritical shocks.

In the ICM, not just relativistic electrons, but also relativistic CR ions (mostly protons) are expected to be present. The ratio of CR to thermal ions, which is constrained by non-detection of gamma-rays from nearby clusters, for instance, in Fermi-LAT observations, seems to be fairly small, less than $10^{-4}-10^{-3}$, averaged over all the cluster volume \citep[e.g.,][]{ryu2019,ha2020,wittor2020}. Yet, we may postulate a scenario that about the same numbers of nonthermal electrons and ions are present in the upstream of merger shocks. However, the presence of nonthermal power-law ions would not change the picture of electron preacceleration described in this paper. In weak $Q_{\perp}$-shocks, unlike suprathermal electrons, higher energy ions are more likely to pass through the shock ramp, rather than to be reflected at the shock potential barrier; hence, the fraction of reflected ions would not increase and the AIC instability would not be enhanced by adding nonthermal power-law ions.
Thus, we conclude that the nonthermal population of CR ions in the upstream plasma would not change the results of this work. Although not shown here, it has been confirmed in PIC simulations we have additionally performed.

A more promising possibility would be the presence of broadband magnetic fluctuations, possibly produced by the turbulence in the ICM. 
Preexisting nonthermal electrons with energies up to $\gamma\sim10^2-10^3$ could be directly injected into DSA via resonant scatterings off those preexisting turbulent waves.
As a matter of fact, previous studies using test-particle simulations for electrons showed that 
electrons could be accelerated via SDA and DSA, regardless of the shock obliquity angle, by interacting with preexisting large-amplitude magnetic fluctuations in interplanetary shocks \citep[e.g.,][]{giacalone2005, guo2015}.
While such acceleration requires efficient wave-particle interactions in the ICM environment, the details of involved processes should be investigated through further simulations.

Additionally, we point that cosmological hydrodynamic simulations showed that radio relics, formed in the turbulent ICM, normally consist of shock surfaces with varying $M_{\rm s}$ and $\theta_{\rm Bn}$ \citep[e.g.,][]{hong2015,roh2019,wittor2021}. 
As suggested by several previous papers including Paper I, it would be interesting to consider a scenario, in which nonthermal electrons, accelerated in the locally supercritical portions of the shock surface, are injected to DSA at subcritical portions.  
Furthermore, the variation of $\theta_{Bn}$ in the shock surface could contribute to the generation of the large-scale modulation in the surface. In the interplanetary shocks, for instance, the surface ripplings on scales larger than the AIC-induced waves have been detected \citep[e.g.,][]{kajdic2019}, which are thought to be generated due to the upstream magnetic field fluctuations produced by the backstreaming ions at $Q_{\parallel}$-portions.

Finally, we thus conjecture that there is still the possibility of electron DSA in radio relics with subcritical shocks, when additional processes and/or ingredients other than those considered in this work are included. 
The investigation of such elements is beyond the scope of this article, so we leave it for future work. 
For now, we conclude that the presence of nonthermal electrons in the ICM ``alone'' would not resolve the issue of electron preacceleration and injection into DSA, and hence could not explain the production of CR electrons in radio relics with subcritical shocks.

\begin{acknowledgments}

This work was supported by the National Research Foundation (NRF) of Korea through grants 2016R1A5A1013277, 2020R1A2C2102800, 2020R1F1A1048189, and 2020R1C1C1012112. J.-H. H. was also supported by the Global PhD Fellowship of the NRF through grant 2017H1A2A1042370. Some of simulations were performed using the high performance computing resources of the UNIST Supercomputing Center.

\end{acknowledgments}

\bibliography{ref_shock}{}
\bibliographystyle{aasjournal}

\end{document}